\documentclass[pra,twocolumn,showpacs,amsmath,amssymb]{revtex4}
\usepackage{graphicx}

\newcommand{\us}{\uparrow}
\newcommand{\ds}{\downarrow}

\newcommand{\ket}[1]{\left| #1 \right\rangle}

\begin {document}
\title {Quantum Entanglement in the S=1/2 Spin Ladder with Ring Exchange}

\author {Jun-Liang Song}
\author {Shi-Jian Gu}
\author {Hai-Qing Lin}

\affiliation{Department of Physics and Institute of Theoretical Physics, The
Chinese University of Hong Kong, Hong Kong, China }

\date{\today}


% 05.70.Jk Critical point phenomena % 75.10.Jm Quantized spin models

\begin{abstract}
In this paper we study the concurrence and the block-block entanglement in the
$S=1/2$ spin ladder with four-spin ring exchange by the exact diagonalization
method of finite cluster of spins. The relationship between the global phase
diagram and the ground-state entanglement is investigated. It is shown that the
block-block entanglement of different block size and geometry manifests richer
information of the system. We find that the extremal point of the two-site
block-block entanglement on the rung locates a transition point exactly due to
$SU(4)$ symmetry at this point. The scaling behavior of the block-block
entanglement is discussed.  Our results suggest that the block-block
entanglement can be used as a convenient marker of quantum phase transition in
some complex spin systems.
\end{abstract}

\maketitle

\section {Introduction}
Entanglement, as one of the most intriguing feature of quantum mechanics
\cite{epr1935}, has become a subject of intense interest in recent years.
Besides being recognized as a kind of crucial resource of quantum computing and
quantum information process \cite{bennett2000,nielsen2000}, it has also
provided new perspectives in problems of various many-body systems.
Particularly, the entanglement can well characterize the features of quantum
phase transition (QPT) \cite{sachdev2000}.  Many works
\cite{osterloh2002,TJOsbornee,SJGuXXZ,unanyan2005,JVidal06,gu2005,SYi06,
XFQian05,vidal2003,korepin2004,SJGUPRL,larsson2005,legeza2006,anfossi2005,
cvenuti2006} have been devoted to understanding the relationship between QPT
and the entanglement in different systems. It has been observed that quantum
phase transitions are signaled by critical behaviors of concurrence
\cite{wootters1998}, a measure of entanglement for two-qubit system, in a
number of spin models
\cite{osterloh2002,TJOsbornee,SJGuXXZ,unanyan2005,JVidal06}. For example, it
was reported that the first derivative of the concurrence diverges at the
transition point in the one-dimensional transverse field Ising model
\cite{osterloh2002}, while the concurrence shows cusp-like behavior around the
critical point in some 2D and 3D spin models \cite{SJGuXXZ}. Besides the
concurrence, the block-block entanglement \cite{audenaert2002} which involves
more system degree of freedom was introduced
\cite{gu2005,vidal2003,korepin2004}.  Especially in fermionic systems in which
the concurrence is usually not applicable, the block-block entanglement can
also manifest interesting properties, such as logarithmic divergence in the
critical region, in a certain class of models \cite{vidal2003,korepin2004}.

However, most of the previous works were restricted within the models with
two-body interaction, the entanglement in the models with three-body or
four-body interactions \cite{bose72} is less investigated and understood. In
fact a system with multi-body interaction is important both in quantum
information theory and condensed matter physics. It was pointed out that a
small cluster of spins with three-body or four-body interactions such as the
four-spin ring exchange could be used for quantum computing \cite{mizel2004,
scarola2004}. Moreover, four-spin ring exchange exists in many physical systems
and plays an important role in understanding the magnetism in several 2D
quantum solids such as solid $^3$He \cite{roger1998} and Wigner Crystals
\cite{okamoto1998}. Therefore, it is of importance to study the properties of
the entanglement in those spin systems with multi-body interactions.

In this paper, we consider a two-legged $S=1/2$ ladder with additional
four-spin ring exchange. The system has a very rich phase diagram
\cite{mueller2002,lauchli2003,hikihara2003,brehmer1999} with many exotic
phases.  We investigate the concurrence and the block-block entanglement in
this system, and try to relate them with the global phase diagram. The rest of
the paper is organized in the following way. In section \ref{sec:model}, we
introduce the model Hamiltonian and its phase diagram.  In section
\ref{sec:conc}, we study the ground-state concurrence and discuss its
relationship with the phase diagram. In section \ref{sec:twosite}, we show that
the two-site block-block entanglement is exactly either maximal or minimal at a
QPT point. In section \ref{sec:be}, we show that the scaling behavior and some
extremal point in the block-block entanglement can be used as marker of QPTs.
In the final section, we summarize our results and draw conclusions.

\section {Model Hamiltonian and phase diagram\label{sec:model} }

\begin{figure}
\includegraphics[width=\columnwidth]{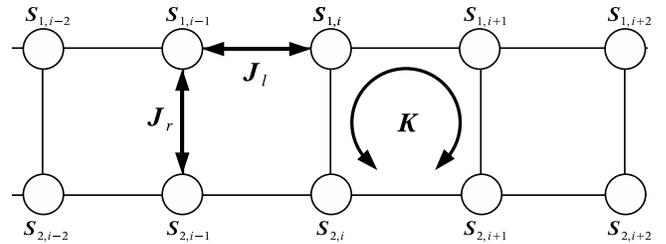}
\caption{A sketch of spin ladder with ring exchange. \label{fig:model}}
\end{figure}

The two-legged $S=1/2$ spin ladder with ring exchange (as shown in Fig.
\ref{fig:model}) is described by the following Hamiltonian
\begin{eqnarray}
{\hat H} &=&J_{r}\sum_{i} {\hat S}_{1,i}{\hat S}_{2,i}
+J_{l}\sum_{i}\left({\hat S}_{1,i}{\hat S}_{1,i+1} + {\hat S}_{2,i}{\hat S}_{2,i+1}\right)\nonumber \\
&+&K \sum_{i}\left( {\hat P}_{i,i+1} + {\hat P}^{-1}_{i,i+1}\right),
\end{eqnarray}
where $i = 1,\dots,N/2$, $N$ is the total number of spins, ${\hat S}_{1,i}$
(${\hat S}_{2,i}$) is 1/2 spin operator on the upper (lower) leg at the $i$th
position, and $J_l$ ($J_r$) is the bilinear exchange constants along the legs
(on the rung) and $K$ is the coupling constant of four-spin cyclic exchange
interaction ${\hat P}$. ${\hat P}_{i,i+1}$(${\hat P}^{-1}_{i,i+1}$) rotates the
four spin in the $i$th plaquette clockwise (counterclockwise), i.e.
\begin{eqnarray}
{\hat P} \left|\begin{array}{ll}
a & b \nonumber \\
d & c \end{array} \right \rangle
=
\left|\begin{array}{ll}
d & a \nonumber \\
c & b \end{array} \right \rangle \textrm{~~~~and~~~~} {\hat P}^{-1}
\left|\begin{array}{ll}
a & b \nonumber \\
d & c \end{array} \right \rangle
=
\left|\begin{array}{ll}
b & c \\
a & d \end{array} \right \rangle,
\end{eqnarray}
and they can be decomposed in terms of spin operator involving bilinear and
bi-quadratic terms,
\begin{eqnarray}
{\hat P}+{\hat P}^{-1}&=&\frac{1}{4} +{\hat S}_a {\hat S}_b+ {\hat S}_b {\hat S}_c+{\hat S}_c {\hat S}_d+{\hat S}_d {\hat S}_a \nonumber\\
&+&{\hat S}_a{\hat S}_c+{\hat S}_b{\hat S}_d \nonumber\\
&+&4\left[ \left({\hat S}_a{\hat S}_b\right)\left({\hat S}_c{\hat S}_d\right)
+\left({\hat S}_a{\hat S}_d\right)\left({\hat S}_b{\hat S}_c\right)\right. \nonumber \\
&-&\left.\left({\hat S}_a{\hat S}_c\right)\left({\hat S}_b{\hat
S}_d\right)\right].
\end{eqnarray}
Following the convention in Ref. \cite{lauchli2003}, we set $J_{l} = J_r =
\cos{\theta}$ and $K=\sin{\theta}$ in the following calculation.

Previous studies \cite{lauchli2003} suggested a rich phase diagram in the
parameter space of $\theta$ shown in the Fig. \ref{fig:conc}. There are
typically six phases (regions): the rung singlet phase, the staggered dimmer
phase, scalar chirality phase, dominant vector chirality region, dominant
collinear spin region and the ferromagnetic phase. Squares in
Fig. \ref{fig:conc} denote first-order phase transitions, circles denote
second-order phase transitions, and the dashed line indicates a crossover
boundary without a phase transition.

Using the exact diagonalization method, we obtain the ground-state concurrence
and the block-block entanglement in a spin ladder up to $N=12\times 2$ sites
with periodical boundary conditions. Although the staggered dimmer and scalar
chirality phases are $Z_2$ symmetry breaking phase with double degeneracy in the
thermodynamic limit, the ground state is unique for most values of $\theta$
except the ferromagnetic phase in a finite-size system. We select $S_{z}=0$
state out of the $N+1$ fold degenerate $S=N/2$ ferromagnetic states in the
following calculation.

\section {Ground-state concurrence\label{sec:conc}}

\begin {figure}
\includegraphics[width=\columnwidth]{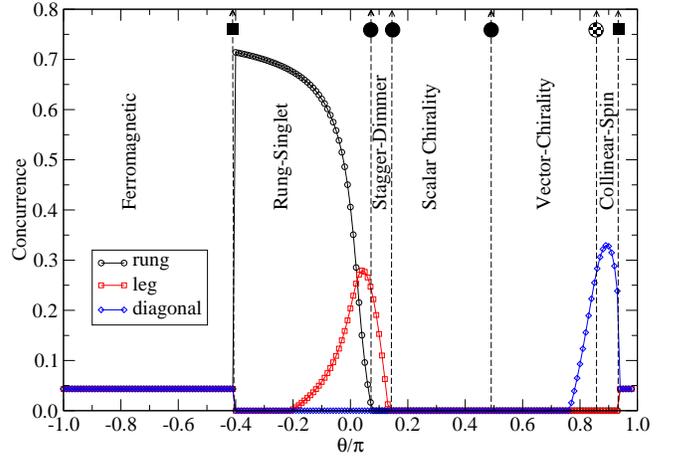}
\caption{(color online) The ground-state concurrence of two spins on a rung,
leg, and diagonal bond as a function of $\theta$ in $N=12\times2$ spin ladder
with ring exchange. The dashed line is the boundary of different phases.
The squares on these line denote a first-order phase transition, the black
circle denotes a second order transistion, while the shaded circle indicates
a transition between two ajacent regions.
\label{fig:conc}}
\end {figure}

The entanglement between the spins at site $i$ and site $j$ can be measured by
the concurrence \cite{wootters1998}. Let $\rho_{ij}$ be the reduced density
matrix which is obtained by tracing out all degrees of freedom of spins except
that at sites $i$ and $j$, and $\tilde{\rho}_{ij}$ be the spin-reversed reduced
density matrix of $\rho_{ij}$, i.e., $\tilde{\rho}= \left(\sigma_y \otimes
\sigma_y \right) \rho^* \left(\sigma_y \otimes \sigma_y \right)$, where
$\sigma_y$ is the Pauli matrix. The concurrence $C$ is given by $C =
\max\left(\lambda_1-\lambda_2-\lambda_3-\lambda_4, 0\right)$, where
\{${\lambda_i}$\} are the square roots of the eigenvalues of the matrix
$\rho\tilde{\rho}$ in descending order. $C = 0$ means no entanglement,
while $C=1$ the maximum entanglement such as that roots in Bell states.

In Fig. \ref{fig:conc} we show the ground-state concurrence as a function of
$\theta$ for a $N=12\times 2$ system. In the ferromagnetic phase, we can see
that the concurrence on any two sites is the same and equals to $1/(N-1)$. It
vanishes in the thermodynamic limit ($N\rightarrow\infty$). In the rung singlet
phase, we can observe a rather large concurrence ($\theta \sim 0.7\pi$) between
the two spins on the same rung. This fact is consistent with the picture that
the ground state is approximated by the product state of spin singlet on the
rungs. Similarly, the concurrence of two adjacent spins on the same leg is
consistent with the physical picture of staggered singlets on the leg in the
staggered dimmer phase. We notice that the peak of concurrence on the leg
($\sim 0.3$) is much smaller than that in the rung singlet phase ($\sim 0.7$).
This is because the ground state in the staggered dimmer phase is two-fold
degenerate in the thermodynamic limit. Then in a finite-size system with
periodic boundary conditions, the ground state is actually a superposition of
these two states, thus the value of the concurrence on the leg reduce to the
half of the original value. In fact if we impose boundary condition in the same
way as that in Ref.  \cite{hakobyan2001}, one of the two degenerate states will
be projected out, the staggered pattern of the leg concurrence appears and the
value on the dimmer leg is nearly $0.6$ which is approaching to $0.7$ in the
rung singlet phase. In both the scalar chirality phase and dominant vector
chirality phase, the concurrence of any pair of spin vanishes. However, at the
cross-over region between the dominant vector chirality and dominant collinear
spin region, an unexpected concurrence on the diagonal pair appears and its
maximal point ($\theta \sim 0.85\pi$) is roughly the crossover point between
the dominant vector chirality region and dominant collinear spin region.

%%%%%%%%%%%%%%%%%%%%%%%%%%%%%%%%%%%%%%%%%%%%%%%%%%%%%

\section {Two-site entanglement of the rung and the $SU(4)$ point
\label{sec:twosite}}

\begin {figure}
\includegraphics[width=\columnwidth]{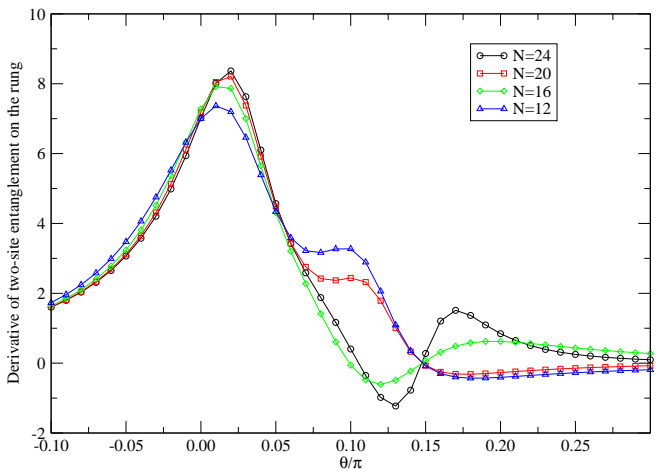}
\caption{(color online) The first derivative of the two-site entanglement on a
rung as a function of $\theta$. Lines of different system sizes cross at the
same point ($\arctan(1/2)$, 0).  \label{fig:deri_rbe}}
\end {figure}

In this section and the following section, we study the block-block entanglement
of various blocks in this system. The block-block entanglement is the von
Neumann entropy $E_v$ of a block of spin in the system. Precisely, it is
calculated as:
\begin{eqnarray}
E_v(A)=-\text{tr}\left(\rho_A\log_2{\rho_A}\right),
\end{eqnarray}
where $A$ is a set of sites and $\rho_A$ is the corresponding reduced density
matrix. If the whole system is in a pure state, then
\begin{eqnarray}
E_v(A)=E_v(B)=-\text{tr}\left(\rho_B\log_2{\rho_B}\right),
\end{eqnarray}
where $B$ is the rest part of the system. Then $E_v(A)$ or $E_v(B)$ describes
how much the block $A$ and the rest of the system are entangled.

Compared to the concurrrence, the block-block entanglement can apply to systems
with much higher degrees of freedom, however, it is only meaningful when the
concerning state is a pure state. In our calculation of finite-size ladders, it
is found that the ground state is always non-degenerate in the region
$-0.40\pi < \theta < 0.95\pi$. Considering the $SU(2)$ symmetry of the
Hamiltonian, the ground state's total spin is also zero in this region.

The term two-site entanglement of a rung means that the von Neumann entropy is
calculated from the reduced density matrix of two spins on the same rung. In
Fig. \ref{fig:deri_rbe} we show the first derivative of the entanglement as a
function of $\theta$ for different system size. From the figure, we observe that
the first derivative of the entanglement on a rung reaches zero exactly at
$\theta=\arctan{(1/2)}\sim 0.148\pi$ which is the QPT point between the
staggered dimmer phase and the scalar chirality phase. We find that this result
is independent of system size. Recently, it was pointed out that, at this QPT
point the system restores $SU(4)$ symmetry\cite{hikihara2003}. Precisely
speaking, at
$\theta_{c}=\arctan(1/2)$, the Hamiltonian commutes with the following operator
\cite{hikihara2003}:
\begin{eqnarray}
{\hat T}=\sum_{i} {\hat S}_{1,i}\cdot {\hat S}_{2,i}.
\end{eqnarray}
We will show that the expectation value of ${\hat T}$ is maximal or minimal
exactly at $\theta = \theta_{c}$ due to the above symmetry.

As discussed above, we can assume the ground state $\ket{\psi_0}$
is non-degenerate around $\theta_c$, which implies $\ket{\psi_0}$
is also the eigenstate the of $\hat{T}$ at $\theta_c$, thus we
have $\hat{T}\ket{\psi_0}=\lambda_t\ket{\psi_0}$ in which
$\lambda_t$ is some real number. Then the first derivative of
$\langle{\hat T}\rangle$ with $\theta$ at $\theta_c$ is
\begin{eqnarray}{\label{eqn:fht}}
\frac{d}{d\theta} \left \langle \psi_0 \right| {\hat T} \left| \psi_0 \right
\rangle &=& \langle \frac{d\psi_0}{d\theta}| {\hat T} | \psi_0 \rangle + \langle
\psi_0 | {\hat T}
| \frac{d\psi_0}{d\theta} \rangle \nonumber \\
&+& \langle \psi_0| \frac{d{\hat T}}{d\theta} | \psi_0 \rangle \\
&=& \lambda_t \langle \frac{d\psi_0}{d\theta} | \psi_0  \rangle +
\lambda_t^*\langle \psi_0 | \frac{d\psi_0}{d\theta} \rangle + 0
\nonumber \\
&=& \lambda_t \frac{d}{d\theta} \left \langle \psi_0 | \psi_0
\right \rangle
\nonumber \\
&=&0 .
\end{eqnarray}

Therefore, the expectation value of ${\hat T}$ reaches local
maximum or minimum at $\theta_c$. Since the ground state has
$(k_x, k_y)=(0, 0)$, the system is invariant under translation
along the leg, so $\langle {\hat S}_{1,i}\cdot {\hat S}_{2,i}
\rangle =\langle {\hat S}_{1,j}\cdot {\hat S}_{2,j} \rangle $
(for any two site $i$ and $j$ in the ladder) is either maximal or
minimal at $\theta_{c}$.

Next we show that there is one-to-one correspondence between the $\langle{\hat
S}_{1,i}\cdot{\hat S}_{2,i}\rangle$ and the two-site entanglement on the rung in
the vicinity of $\theta_c$. Let $\rho_{ij}$ be the reduced density matrix of
spins of sites $i$ and $j$ of the ground state. In the basis
\{$\ket{\ds\ds},\ket{\ds\us}, \ket{\us\ds},\ket{\us\us}$\},
$\rho_{ij}$ has the following form due to the $U(1)$ symmetry of the ground
state in the concerning region($0.1\pi < \theta < 0.2\pi$):
\begin{eqnarray}
\rho_{ij} = \left( \begin{array}{cccc}
u^+ & 0 & 0 & 0 \\
0 & w_1 & z^* & 0 \\
0 & z & w_2 & 0 \\
0 & 0 & 0 & u^-
\end{array} \right).
\end{eqnarray}
Moreover, in the vicinity of $\theta=\theta_c$, the ground state is
unique and its total spin $S=0$, which implies the ground state is also
invariant under any rotations. Particularly, $\rho_{ij}$ is invariant under
the rotation around the $x$ axis:
\begin{eqnarray}\label{eqn:com}
\left[ \sigma^{x}_{i}+\sigma^{x}_{j}, \rho_{ij}\right] = 0 .
\end{eqnarray}
From Eq. (\ref{eqn:com}), and the condition $\text{tr}\rho=1$, we have
\begin{equation}
u^+=u^-=\frac{1+2z}{4},\;\; w_1=w_2=\frac{1-2z}{4},
\end{equation}
\begin{equation}
z = z^*,\;\;\; \left\langle{\hat S}_{1,i}\cdot{\hat S}_{2,i}\right\rangle=
-\frac{3z}{2},
\end{equation}
Then the block-block entanglment on the rung $E_r$ is
\begin{equation}
E_{r}=-3u^+\log_2{u^+}-(w_1-z)\log_2(w_1-z) .
\end{equation}
From the above equations it is clear that this extremal behavior of two-site
entanglement on the rung is directly related to the $SU(4)$ symmetry.

\section {scaling behavior of the block-block entanglement\label{sec:be}}

\begin{figure}
\centering
\includegraphics[width=\columnwidth]{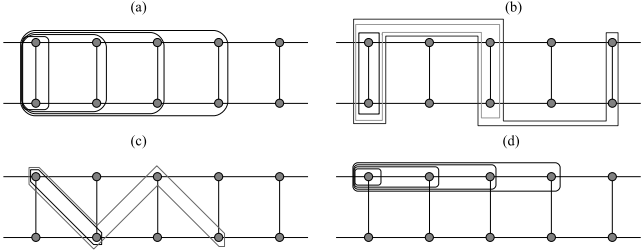}
\caption{Four choices of increasing the block size: (a) is the single block,
(b) is the stripe block, (c) is the zigzag block, (d) is one-leg block.
\label{fig:blocx}}
\end{figure}

\begin{figure*}
\includegraphics[width=\textwidth]{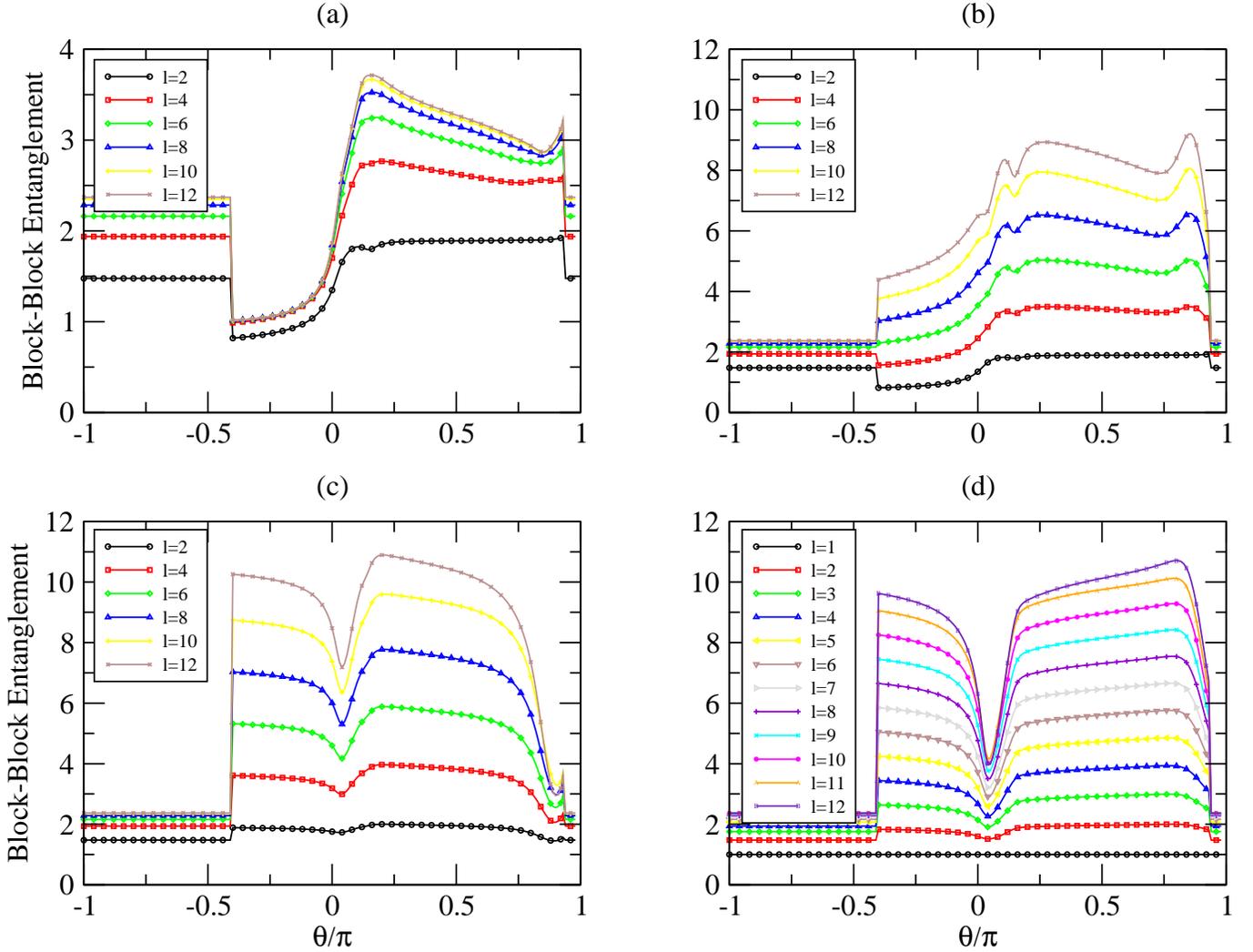}
\caption{(color online) The block-block entanglement of different block size
$l$ as a function of $\theta$ in $N=12\times2$ ladder. The geometry of the
blocks in (a)(b)(c)(d) are specified in Fig. \ref{fig:blocx}(a)(b)(c)(d). }
\label{fig:blockscaling}
\end{figure*}

In this section, we focus on the scaling behavior of the
block-block entanglement, namely how the block-block entanglement
behaves as the block size changes. Unlike the case in the one
dimensional chain, the ladder geometry has provided us many
choices of how to select the block's shape and how to increase
the block size. As shown in Fig. \ref{fig:blocx}, we choose four
different ways to increase the block size.

First, we notice that in the ferromagnetic phase, the value of the block-block
entanglement depends only on the size of the block. It is independent of the
block geometry. This is because any two sites in this state is essentially
equivalent as we have already seen in Sec. \ref{sec:conc}. In fact an explicit
expression of block-block entanglement as a function of block size $l$ can
be obtained in the following calculation.

The ferromagnetic $S^z_{tot}=\sum_i{S^z_i}=0$ state
$\psi_{FM}(S^z_{tot}=0)$ could be obtained by applying lowering operator
on the $\psi_{FM}(S^z_{tot}=N/2)$ $N/2$ times.

\begin{eqnarray}
\left|\phi_{FM}(0)\right\rangle =
\frac{\left(\sum_{i=1}^{N}{\hat S}_i^-\right)^{\frac{N}{2}}}
{\sqrt{\left(\frac{N}{2}\right)!}}
\underbrace{\left|\uparrow\uparrow\uparrow\ldots\uparrow\right\rangle}_{N},
\end{eqnarray}

where ${\hat S}_i^-$ is the spin lowering operator at site $i$ and $N$ is the
total number of sites. In the basis \{$\ket{\ds\ds},\ket{\ds\us},
\ket{\us\ds},\ket{\us\us}$\}, all the coefficients of
$\psi_{FM}\left(S^z_{tot}=0\right)$ are the same: $\sqrt{{(N/2)!(N/2)!}/{N!}}$.
The matrix element $\rho_l(p,q)$ could be obtained explicitly:

\begin{eqnarray}
\rho_l(p,q) = \left \{ \begin{array}{ll} \frac {
\left(\frac{N-l}{2}\right)!  \left(\frac{N}{2}\right)!
\left(\frac{N}{2}\right)!  }{ \left(\frac{N-l}{2}-p_z\right)!
\left(\frac{N-l}{2}+p_z\right)!  \left(N\right)!  }
&\textrm{if $p_z = q_z$}\\
0&\textrm{if $p_z \neq q_z$}
\end{array} \right ..
\end{eqnarray}

In the above expression, $\rho_l$ is reduced density matrix of the block
consisting of $l$ spins(we assume $l \leq N/2$), $p$ and $q$ are the column and
row index of $\rho_l$, $p_z$ and $q_z$ are the corresponding $S^z_{tot}$
number.

After diagonalizing this matrix, there are only $l+1$ nonzero eigenvalues
$\lambda_{p_z}$ with $p_z=-l/2, -l/2+1, \dots, l/2$. Then the block-block
entanglement can be obtained as:
\begin{eqnarray}
\lambda_{p_z}&=&
\frac{\left(l\right)! \left(\frac{N-l}{2}\right)!
\left(\frac{N}{2}\right)! \left(\frac{N}{2}\right)! }{
\left(\frac{l}{2}-p_z\right)! \left(\frac{l}{2}+p_z\right)!
\left(\frac{N-l}{2}-p_z\right)! \left(\frac{N-l}{2}+p_z\right)! \left(N\right)!
}, \nonumber \\
E_v({l}) &=& \sum_{p_z=-l/2}^{l/2}{-\lambda_{p_z}\log_2{\lambda_{p_z}}}.
\label{eqn:vnl}
\end {eqnarray}

When $l$ and $N$ are large, the summation in Eq. (\ref{eqn:vnl})
can be replaced by an integral, and the function $\lambda_{p_z}$
can be approximated by the Gaussian Distribution. Thus we can
approximate $E_v{l}$ by the following expression:

\begin{eqnarray}
E_v(l) \sim -\frac{1}{2}\log_{2} {\left(\frac{1}{l}+\frac{1}{N-l}\right)}
+\frac{1}{2}\log_{2}{\left(\frac{\pi e}{2}\right)},
\end{eqnarray}
which suggests that $E_v(l)$ diverges logarithmically as the size
$l$ increases.

Secondly, in Fig. \ref{fig:blockscaling}(a), in the most region of rung singlet
phase, the block-block entanglement converges to some finite value quickly,
while in \ref{fig:blockscaling}(b)--(d), it is almost proportional to the
block size. Since the ground state is approximately the rung singlet product
state, the block-block entanglement is proportional to the number of bonds
cross the boundary of the block such as in case (c) and (d). In case (b), the
situation is different since the number of bonds cross the boundary of block is
finite, thus the short-ranged correlations between the rungs plays important
role in (b) and explains this proportionality.

Next, in Fig. \ref{fig:blockscaling}(b)--(d), we find that some local extreme
points of the block-block entanglement may be the QPT points. Previous
studies\cite{lauchli2003, hikihara2003} suggested that the transition point
between the rung singlet and staggered dimmer to be $0.06 \pi \sim 0.08 \pi$
which is quite near the local maximum point $0.07\pi$ in (b), local minimum
$0.05\pi$ in (c) and (d). The transition point between the staggered dimmer and
scalar chirality phase is exactly $0.148\pi$ which is also near one minimal
point $0.14\pi$ in (b). As for the crossover point between the dominant vector
chirality region and the dominant collinear spin region, (b)-(d) all suggest
the value $\sim 0.85\pi$ which is coincidence with the value obtained in
previous works\cite{lauchli2003}.

In general, the scaling behavior of the above four choices of blocks could be
categorized into two kinds, Fig. \ref{fig:blockscaling}(a) and (b)--(d). In (a),
the number of ladder bonds across the boundary between two blocks is a finite
value which equals to $4$ independent of the block size, while in (b)--(d),
the number is proportional to the block size $l$. In the latter case, the
short-ranged correlation across the boundary bonds is main contribution to the
block-block entanglement, thus the block-block entanglement is always
proportional to the size of the block as we have seen in Fig. \ref{fig:blocx}.
In the former case, the main contribution to the block-block entanglement comes
from the long-range correlation between the sites in the block and the sites
outside the blocks. It is expected that around the QPT point, in the former
case the scaling behavior changes abruptly, e.g. from convergence to finite
value to divergence logarithmically, while in the latter case, there may exist
extremal point of the block-block entanglement which is an indication of
QPT.

\section {summary and acknowledgment}

In summary, we have studied the concurrence and the block-block
entanglement in the ground state of the $S=1/2$ spin ladder with
ring exchange. In both the rung singlet and staggered dimer
phase, the behaviors of the ground-state concurrence are
consistent with the corresponding dominant configurations. The
extremal point of the two-site block-block entanglement coincide
with the QPT point due to the $SU(4)$ symmetry, and such a
symmetry is obviously independent of the system size. We have also
investigated the scaling behavior of the block-block entanglement
for different block geometry and block size. We have identified
three kinds of typical scaling behavior in this model, namely
increasing the size of block, the block-block entanglement (a)
converges to some finite value, (b) diverges logarithmically with
size, (c) diverges proportional with the size. However, as we can
see that, there's no signature of the QPT between the scalar
chirality phase and dominant vector chirality phase.

This work is supported by the Earmarked Grant for Research from the Research
Grants Council of HKSAR, China (Project CUHK N\_CUHK204/05 and HKU\_3/05C).


\begin{references}


% Original work on EPR
\bibitem{epr1935}
A. Einstein and B. Podolsky and N. Rosen, Phys. Rev. {\bf 47}, 777 (1935).


% Review article of entanglement in QI theory
\bibitem{bennett2000}
C. H. Bennett and D. P. Divincenzo, Nature {\bf 404}, 247 (2000).


\bibitem{nielsen2000}
M. A. Nielsen and I. L. Chuang, {\it Quantum Computation and Quantum
Information}, (Cambridge University Press, 2000).


\bibitem{sachdev2000}
S. Sachdev, {\it Quantum Phase Transitions}, (Cambridge University Press, 2000).


% Entanglement in spin systems:
\bibitem{osterloh2002}
A. Osterloh, Luigi Amico, G. Falci, Rosario Fazio,  Nature {\bf 416}, 608
(2002).

\bibitem{TJOsbornee}
T. J. Osborne and M.A. Nielsen, Phys. Rev. A {\bf 66}, 032110(2002).

\bibitem{SJGuXXZ}
S. J. Gu, H. Q. Lin, and Y. Q. Li, Phys. Rev. A {\bf 68}, 042330 (2003); S. J.
Gu, G. S. Tian, H. Q. Lin,  Phys. Rev. A {\bf 71}, 052322 (2005).

% Many-particle entanglement in the gaped antiferromagnetic Lipkin model
\bibitem{unanyan2005}
R. G. Unanyan and C. Ionescu and M. Fleischhauer, Phys. Rev. A {\bf 72}, 022326
(2005).


\bibitem{JVidal06}
J. Vidal, quant-ph/0603108.


\bibitem{SYi06}
S. Yi and H. Pu, Phys. Rev. A {\bf 73}, 023602 (2006).

\bibitem{XFQian05}
X. F. Qian, T. Shi, Y. Li, Z. Song, and C. P. Sun, Phys. Rev. A {\bf 72},
012333 (2005).

% Another arguement using operator
\bibitem{gu2005}
S. J. Gu and G. S. Tian and H. Q. Lin, New J. Phys. {\bf 8}, 61 (2006).


% Block-block-block entanglement scaling in Ising model
\bibitem{vidal2003}
G. Vidal, J. I. Latorre, E. Rico, and A. Kitaev, Phys. Rev. Lett. {\bf 90},
227902 (2003).


\bibitem{korepin2004}
V. E. Korepin, Phys. Rev. Lett. {\bf 92}, 096402 (2004).


% Entanglement in fermionic systems:
\bibitem{SJGUPRL}
S. J. Gu, S. S. Deng, Y. Q. Li, H. Q. Lin, Phys. Rev. Lett. {\bf 93}, 086402
(2004); S. S. Deng, S. J. Gu, and H. Q. Lin, Phys. Rev. B in press.

% Entanglement Scaling in the One-Dimensional Hubbard Model at Criticality
\bibitem{larsson2005}
Daniel Larsson and Henrik Johannesson, Phys. Rev. Lett {\bf 95}, 196406 (2005).

\bibitem{legeza2006}
{\"O}. Legeza and J. S{\'o}lyom, Phys. Rev. Lett. {\bf 96}, 116401 (2006).

\bibitem{anfossi2005}
Alberto Anfossi, Paolo Giorda, Arianna Montorsi, and Fabio  Traversa, Phys. Rev.
Lett. {\bf 95}, 056402 (2005); Alberto Anfossi, Cristian Degli Esposti Boschi,
Arianna Montorsi, and  Fabio Ortolani, Phys. Rev. B  {\bf 73}, 085113 (2006).

\bibitem{cvenuti2006}
L. {Campos Venuti}, C. {Degli Esposti Boschi}, M. Roncaglia, and  A. Scaramucci,
Phys. Rev. A {\bf 73}, 010303(R) (2006).


% Definition of concurrence
\bibitem{wootters1998}
W. K. Wootters, Phys. Rev. Lett. {\bf 80}, 2245 (1998).


% Block-block entanglement
\bibitem{audenaert2002}
K. Audenaert, J. Eisert, M. B. Plenio, and R. F. Werner, Phys. Rev. A {\bf 66},
 042327 (2002).


% Thermal entanglement properties of small spin clusters
\bibitem{bose72}
Indrani Bose and Amit Tribedi, Phys. Rev. A {\bf 72}, 022314 (2005).


% Three- and Four-Body Interactions in spin based quantum computers
\bibitem{mizel2004}
Ari Mizel and Daniel A. Lidar, Phys. Rev. Lett. {\bf 92}, 077903 (2004).

% Chirality in Quantum Computation with spin clusters Qubits
\bibitem{scarola2004}
V. W. Scarola and K. Park and S. Das Sarma, Phys. Rev. Lett. {\bf 93}, 120503
(2004).


% Ring exchange in two dimensional solid Helium_3
\bibitem{roger1998}
M. Roger and C. B{\"{a}}uerle and Yu. M. Bunkov and A. S. Chen and\ H. Godfrin,
Phys. Rev. Lett. {\bf 80}, 1308 (1998).

% Wigner crystal
\bibitem{okamoto1998}
Tohru Okamoto and Shinji Kawaji, Phys. Rev. B {\bf 57}, 9097 (1998).


% Four spin ring exchange model details, P4+P4-1 in terms of spin operators !
\bibitem{mueller2002}
M. M{\"{u}}eller and T. Vekua and H. J. Mikeska, Phys. Rev. B, {\bf 66}, 134423
(2002).

% Phase diagram of ring exchange model ! DMRG method
\bibitem{lauchli2003}
A. L{\"{a}}uchli and G. Schmid and M. Troyer, Phys. Rev. B {\bf 67}, 100409(R)
(2003).


% Spin-Chirality Duality in a Spin Ladder with Four-Spin Ring Exchange
\bibitem{hikihara2003}
Toshiya Hikihara and Tsutomu Momoi and Xiao Hu, Phys. Rev. Lett. {\bf 90},
087204 (2003).


% Phase diagram obtained by perturbation method, exact diagonalization and
% exact solutions.
\bibitem{brehmer1999}
S. Brehmer and H. J. Mikeska and  M. M{\"{u}}ller and N. Nagaosa and S. Uchida,
Phys. Rev. B. {\bf 60}, 329 (1999).


% Open ladder with one spin attached at each side.
\bibitem{hakobyan2001}
T. Hakobyan and J. H. Hetherington and M. Roger,  Phys. Rev. B, {\bf 63}, 144433
(2001).

\end{references}
\end {document}